\documentclass[letterpaper, 10 pt, conference]{ieeeconf}  

\IEEEoverridecommandlockouts                              
\usepackage{graphics} 
\usepackage{epsfig} 
\usepackage{times} 
\usepackage{amsmath} 
\usepackage{amssymb}  
\usepackage[ruled,vlined]{algorithm2e}
\usepackage{xcolor}
\usepackage{xcolor,colortbl}
\usepackage{booktabs}
\usepackage{flushend}

\let\labelindent\relax
\usepackage[inline]{enumitem}
\definecolor{brickred}{rgb}{0.8, 0.25, 0.33}

\usepackage[style=ieee, hyperref=false]{biblatex}
\AtEveryBibitem{
    \clearfield{url}
    \clearfield{urlyear}
    \clearfield{urlmonth}
    \clearfield{doi}
    \clearfield{issn}
    \clearfield{eprint}
}
\usepackage{siunitx}
\usepackage{subcaption}
\AtBeginBibliography{\footnotesize}

\title{\LARGE \bf
An Efficient Risk-aware Branch MPC for Automated Driving that is Robust to Uncertain Vehicle Behaviors}

\author{Luyao Zhang$^{1}$, George Pantazis$^{1}$, Shaohang Han$^{2}$ and Sergio Grammatico$^{1}$
\thanks{*This work is partially supported by NWO under project AMADeuS and by the ERC under project COSMOS.}
\thanks{$^{1}$Luyao Zhang, George Pantazis and Sergio Grammatico are with the Delft Center for Systems and Control, Delft University of Technology, The Netherlands.
        {\tt\small \{l.zhang-7, 
        g.pantazis, 
        s.grammatico\}@tudelft.nl.}
        }%
\thanks{$^{2}$Shaohang Han is with the
Division of Robotics, Perception and Learning, KTH
Royal Institute of Technology, Sweden. {\tt\small shaohang@kth.se}.}%
}

\begin{document}

\maketitle
\thispagestyle{empty}
\pagestyle{empty}

\begin{abstract}
One of the critical challenges in automated driving is ensuring safety of automated vehicles despite the unknown behavior of the other vehicles.
Although motion prediction modules are able to generate a probability distribution associated with various behavior modes, their probabilistic estimates are often inaccurate, thus leading to a possibly unsafe trajectory.
To overcome this challenge, we propose a risk-aware motion planning framework that appropriately accounts for the ambiguity in the estimated probability distribution.
We formulate the risk-aware motion planning problem as a min-max optimization problem and develop an efficient iterative method by incorporating a regularization term in the probability update step.
Via extensive numerical studies, we validate the convergence of our method
and demonstrate its advantages compared to the state-of-the-art approaches. 

\end{abstract}


\section{Introduction}
After two decades of development, automated vehicles have been able to navigate a variety of traffic scenarios successfully.
Nevertheless, operating in highly interactive environments, such as intersections without traffic lights, still remains a significant challenge. 
One main reason is that an automated vehicle struggles to properly account for the different behavior modes of the surrounding vehicles.
Specifically, in the classic planning framework \cite{fan_baidu_2018}, the motion planner receives the most likely predicted trajectories of the surrounding vehicles, based on a motion prediction module, and then generates a motion plan without considering multi-modal behaviors.
One drawback of not considering multiple behaviors of the surrounding vehicles is that the generated motion plan might be overly aggressive or overly conservative, thus resulting in collisions or traffic jams, respectively. 

\begin{figure}[!t]
    \centering
    \includegraphics[width=1.0\columnwidth]{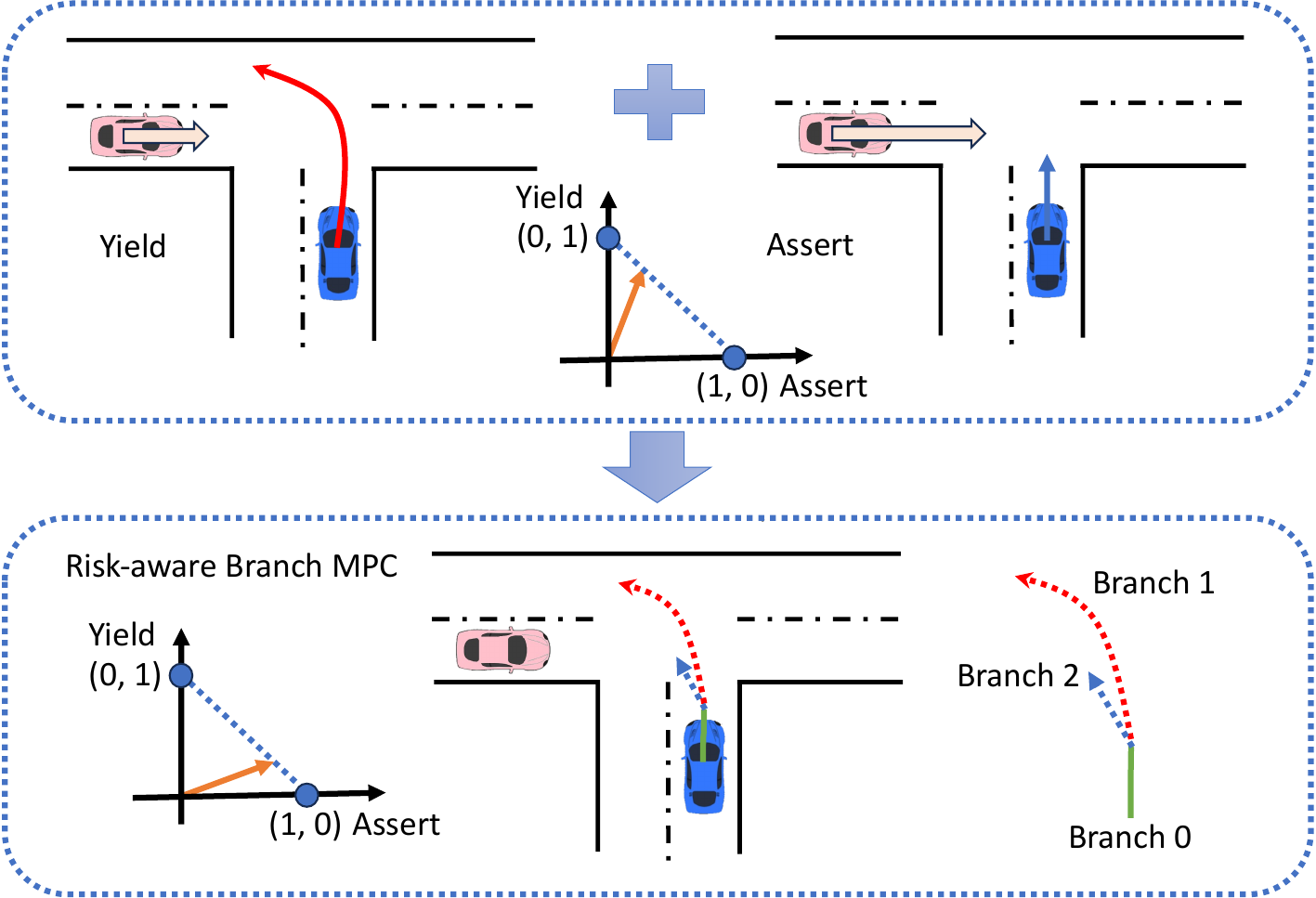}
    \caption{
    Unsignalized intersection-crossing scenario. The other vehicle (in pink) has two potential behavior modes: ``Yield" and ``Assert''. In this case study, the motion predictor assesses the likelihood of each behavior mode and indicates that the other vehicle is more likely to ``Yield". However, the behavior mode ``Assert" can result in a potential collision.  
    To avoid unsafe motion, the risk-aware branch MPC planner generates a trajectory tree that considers the different behavior modes by taking into account their associated ambiguity.
In this example, it focuses more on the behavior mode ``Assert'' by appropriately reshaping the probability distribution.} 
    \label{fig:intersection_crossing}
\end{figure}

To properly consider multi-modal behaviors, the so-called branch model predictive control (BMPC) \cite{ChenMultiModal2022} has been proposed, also referred to as contingency planning in \cite{HardyContingency2013, AlsterdaContingency2019}, or trajectory tree motion planning in \cite{huActiveUncertaintyReduction2023a, WangInteraction2023, BatkovicScenario2021}. 
An advantage of BMPC planners is their ability to utilize the multi-modal trajectory prediction provided by the latest motion prediction module \cite{guDenseTNTEndtoendTrajectory2021}. 
In contrast to traditional robust motion planners, which aim to generate a motion plan that accommodates all predicted trajectories, BMPC planners construct a trajectory tree with multiple branches corresponding to (possibly) different behavior modes. 
Thus, BMPC planners can avoid generating overly conservative motion plans since only the shared branch needs to adapt to all potential predicted scenarios. 

However, a significant challenge arising with multi-modal behaviors is that estimates on the probability of each tree branch might be inaccurate \cite{ChenJointPlanning2024}. 
To address this issue, it is necessary to deal with ambiguity in the probability estimates by leveraging tools from risk-aware stochastic optimization.  
In \cite{ChenMultiModal2022, SopasakisRiskAverse2019, LiMultipolicy2023}, the authors propose a risk-aware MPC framework, where they employ the so-called Conditional Value at Risk (CVaR), a risk measure that accounts for unlikely scenarios at the tail of the probability distribution. 
They then apply this framework to risk-aware motion planning, leveraging the dual form of CVaR \cite{ShapiroSP2021} to recast the original problem as a min-max optimization problem with a nonconvex-concave structure. \par 
The works \cite{ChenMultiModal2022, SopasakisRiskAverse2019} address the min-max reformulation by converting the inner maximization problem into a minimization problem via the dualization technique. However, such a formulation introduces additional auxiliary decision variables and hard constraints. 
Furthermore, the nonconvexity of such constraints due to the collision cost renders this problem computationally challenging to solve. 
Another popular category of algorithms for min-max optimization would solve the maximization problem and perform a gradient descent step for the minimization problem. 
Particularly, in \cite{nouiehed_solving_2019}, the maximization problem is solved approximately by employing multiple gradient ascent steps.
In contrast, in single-loop algorithms such as the gradient descent-ascent (GDA), the update for one variable only occurs once before another variable is updated. 
However, the GDA method fails to converge even for some simple bilinear problems \cite{RazaviyaynMinmax2020}.
To improve convergence, Lu et al. \cite{LuHBSiA2020} propose the so-called Hybrid Block Successive Approximation (HiBSA), where a regularization term is introduced to the maximization problem, resulting in a perturbed version of the original projected gradient ascent step.

To iteratively solve the outer minimization problem, a projected gradient descent (PGD) method is employed in \cite{nouiehed_solving_2019}.
However, as the feasible set in the motion planning problem is often non-convex, the projection to such a set is computationally challenging. 
To compute the update more efficiently, Li et al. \cite{LiMultipolicy2023} replace the gradient descent with iterative LQR (iLQR), an efficient numerical
optimization method that exploits the sparse structure inherent in optimal control problems \cite[Chapter 8]{rawlings_model_2017}.
iLQR is initially applied to unconstrained motion planning in \cite{Li_iLQR_2004}, while in \cite{TassaControliLQR2014}, a projected quasi-Newton method is employed to consider input constraints. 
To handle more general inequality constraints, a barrier function and an augmented Lagrangian function are integrated into the framework of iLQR in \cite{ChenConiLQR2019} and \cite{HowellALTRO2019}, respectively.
The authors in \cite{LiMultipolicy2023, DaComprehensive2022} have recently developed iLQR-based solvers for the BMPC problem.

In this work, we propose an efficient implementation of a risk-aware motion planner for applications in autonomous driving. Our contributions with respect to the related literature can be summarized as follows: 

\begin{enumerate}[label=(\roman*)]
\item 

To circumvent the computational challenges of the formulations in \cite{ChenMultiModal2022, SopasakisRiskAverse2019}, our work builds upon \cite{LiMultipolicy2023} and proposes an iterative algorithm based on the augmented Lagrangian iterative linear quadratic regulator (AL-iLQR) to efficiently solve the minimization problem in the min-max reformulation of the risk-aware motion planning problem. 

\item We combine the method above with a variant of  Hybrid Block Successive Approximation (HiBSA) \cite{LuHBSiA2020}; specifically, to address scenarios of oscillatory behavior observed by implementing the numerical method in \cite{LiMultipolicy2023}, we introduce a regularization term to the inner maximization problem. 
We empirically validate the effectiveness of such an additional term via extensive numerical simulations. 

\item We implement our method in C++ and
demonstrate its applicability in real-time motion planning by considering different case studies of unsignalized\footnote{The term ``unsignalized'' is used in the literature to refer to the absence of traffic lights.} intersection crossing. Even though this work focuses mainly on unsignalized intersections, our proposed algorithm can be applied to a variety of traffic scenarios. 
\end{enumerate}
This paper can be seen as a preliminary result towards a more challenging research goal:
integrating a risk-aware motion planner together with a behavior planner, and applying them to highly interactive automated driving scenarios \cite{zhang_automated_2024, liu2022interaction}. \par 

\section{Problem formulation}
 \subsection{Unsignalized intersection crossing} 
We consider an unsignalized intersection-crossing scenario, where the ego vehicle interacts with the surrounding vehicles.
Fig. 1 illustrates such an example, where the ego vehicle makes a left turn while the other vehicle follows a straight path. 
To turn left successfully, the ego vehicle should take into account the different behavior modes of the other vehicle.
In this example, two behavior modes can be identified for the other vehicle. 
The first behavior mode is called ``Yield" mode, i.e., the other vehicle decides to slow down and wait for the ego vehicle to turn left. 
In this case, the ego vehicle can maintain the current speed (assuming it is sufficiently slow to make a left turn) and proceed through the intersection. 
The second behavior mode is the ``Assert" mode, where the other vehicle does not decelerate and continues following the straight path. 
Consequently, the ego vehicle should slow down to avoid a potential collision. 

The traffic scenario in Fig. \ref{fig:intersection_crossing} illustrates how different behavior modes can have a significant effect on motion planning. 
To account for the presence of multiple behavior modes of surrounding vehicles, we introduce a motion planner based on the branch MPC framework. 

\subsection{Nominal Branch MPC}

A branch MPC planner generates a so-called trajectory tree, i.e.,  a decision tree whose branches correspond to distinct behavior modes of the surrounding vehicles. 
An example of the trajectory tree is illustrated in Fig. \ref{fig:intersection_crossing}, which comprises one shared branch and a branching point that leads to two possible outcomes.
We denote the time step when the shared branch splits into multiple individual ones by $T_s$ and the length of the entire planning horizon by $T$.

In the trajectory tree, the initial control inputs are constrained to remain consistent over the interval $[0, T_s-1]$ to allow for adaptation to all potential predicted scenarios.  
We denote the tree of control inputs by $\bar{\mathbf{u}} := \left( \bar{\mathbf{u}}^0, \bar{\mathbf{u}}^i \right)_{i\in [1, d]}$, where $d$ is the number of branches,
$\bar{\mathbf{u}}^0 := (u^0_{t})_{t\in[0, T_{s}-1]}$ represents the shared control input sequence from time step $0$ to $T_s-1$, and $\bar{\mathbf{u}}^i := (u_t^i)_{t\in[T_{s}, T-1]}$ denotes the control input sequence for branch $i$ from time step $T_s$ to $T-1$. 
We consider a state $x^i_t \in \mathbb{R}^{n_x}$ corresponding to branch $i$ at time step $t$, which evolves according to the dynamics $x^i_{t+1}=f(x_t^i, u_t^i)$, where $f: \mathbb{R}^{n_x} \times  \mathbb{R}^{n_u} \rightarrow  \mathbb{R}^{n_x}$ is a nonlinear function. The initial state for the shared branch is denoted by $x_0$, while $x_{T_s}$ denotes the state at the start of the branching.  
Note that the state evolution can be expressed as a function of the initial condition $x_0$ and the input $\bar{\mathbf{u}}$.
We define the cost function $J: \mathbb{R}^{n_x} \times \mathbb{R}^{n_u T_s + n_u (T - T_s) d} \rightarrow \mathbb{R}$  of the branch MPC as the weighted sum of the branch costs:
\begin{multline}
    J(x_0,\bar{\mathbf{u}}) := \underbrace{\sum_{t=0}^{T_s-1} \ell^0(x^0_t, u^0_t)}_{J^0(x_0, \bar{\mathbf{u}}^0)} \\ +  
    \sum_{i=1}^d p^i \underbrace{
    \left( \sum_{t=T_s} ^{T-1} \ell^i(x^i_t, u^i_t) + \ell_T^i(x^i_T) \right)}_{J^i(x_{T_s}, \bar{\mathbf{u}}^i)}, \label{eq:bmpc_expection}
\end{multline}
where $\ell^i: \mathbb{R}^{n_x} \times \mathbb{R}^{n_u} \rightarrow \mathbb{R}$ and $\ell^i_T: \mathbb{R}^{n_x} \rightarrow \mathbb{R}$ denote the stage and final cost, respectively, and $p=(p^i)_{i \in [1,d]}$ is the collection of probability estimates for all branches, which usually originates from an upstream prediction module or behavior planner and takes values in the simplex $P=\{p \in \mathbb{R}^d_{\geq 0} \mid \sum_{i=1}^d p^i=1\}$.
In practice, however, the probability estimate can often be inaccurate, resulting in possibly unsafe motion plans. \par

\subsection{Risk-aware branch MPC}
Revisiting the unsignalized intersection-crossing problem in Fig. 1, the motion predictor assesses the likelihood of each behavior mode and indicates that mode ``Yield" has a larger probability of occurrence.
However, the behavior mode ``Assert" could lead to a potential collision with the ego vehicle, which implies that ambiguity in the probability estimates carries a significant risk for the safety of both vehicles. 
Therefore, the ego vehicle should be more aware of the risk associated with potential misinterpretation of the other vehicle's intentions.
To generate a robust motion plan for such intricate cases, we develop a risk-aware branch MPC planner, which focuses more on the potentially dangerous scenarios by minimizing a risk measure of the cost function.

\subsubsection{Preliminaries}
Let us first introduce the concept of risk measures. 
A popular class of risk measures used in stochastic optimization is the so-called coherent risk measures, which satisfy certain properties, including convexity, monotonicity, translation equivariance, and positive homogeneity; see \cite{ShapiroSP2021} for technical details.
While widely used as a coherent risk measure, the expectation fails to effectively account for events that lie at the tail of a given distribution.
To circumvent this issue,  Conditional Value at Risk (CVaR) \cite{ShapiroSP2021, rockafellar_deviation_2002} is a popular means for improved risk assessment.
For computational efficiency, the dual representation of a coherent risk measure \cite[Eq. 6.40]{ShapiroSP2021} is commonly employed:
\begin{align}
    \rho(X) := \sup_{\mathbb{Q} \in \mathcal{A}} \; \mathbb{E}_{\mathbb{Q}}[X], \label{eq:CVaR_dual}
\end{align}
where $X$ is a random variable and the ambiguity set $\mathcal{A}$ is closed and convex. 
The dual representation indicates that a coherent risk measure can be viewed as the worst-case expectation with respect to all probability distributions in the ambiguity set. 

In this work, we consider that uncertainty arises from the unknown discrete behavior modes of the surrounding vehicles.
This motivates the study of discrete probability distributions $q=(q^i)_{i \in [1,d]}$, where
the considered ambiguity set of $\text{CVaR}_{\alpha}$ for $\alpha \in [0, 1]$, is the intersection of the probability simplex and the set of boxes, based on the nominal probability vector $p$:
\begin{align}
\mathcal{A}_\alpha(p) = \left\{ q \in \mathbb{R}^d \mid \sum_{i=1}^{d} q^i = 1,\; q^i \geq 0,\; \alpha q^i \leq p^i \right\}. \label{eq:ambiguity_set}
\end{align}
Selecting $\alpha = 0$ implies a lack of confidence in the nominal probability distribution, resulting in the ambiguity set being equivalent to the entire probability simplex. 
For $\alpha = 1$, the decision maker has more confidence in the nominal probability distribution.

\subsubsection{Risk-aware formulation} 
By adopting CVaR as a risk measure and leveraging its dual form \eqref{eq:CVaR_dual}, the original cost in \eqref{eq:bmpc_expection} has the following risk-aware counterpart:
\begin{align}
    J_{\text{R}}(x_0, \bar{\mathbf{u}}) := J^0(x_0, \bar{\mathbf{u}}^0) + 
    \max_{q \in \mathcal{A}_\alpha(p)}\; \sum_{i=1}^d q^i J^i(x_{T_s}, \bar{\mathbf{u}}^i), \label{eq:CVaR_cost}
\end{align}
where the ambiguity set $\mathcal{A}_\alpha(p)$ is given in \eqref{eq:ambiguity_set}. Informally speaking, the solution to this maximization problem tends to assign a larger probability to the branch with a higher cost.
With the cost function in (\ref{eq:CVaR_cost}), we can formulate the risk-aware motion planning problem (RAMP) as follows:
\begin{equation}
(\text{RAMP}) \left\{
    \begin{aligned}
        \min_{\bar{\mathbf{u}}}\; &J_{\text{R}}(x_0, \bar{\mathbf{u}}) \\
        \text{s.t. }\; &x^0_{t+1} \!\!= \!\!f(x^0_t, u^0_t),\;\forall t \in [0, T_s-1],  \\
        &h^0(x_t^0, u_t^0) \leq 0,\,  \forall t \in [0, T_s-1],\;  \\ 
        & x^i_{T_s}=x^j_{T_s},\; \forall i,j \in [0,d], \\
        &x^i_{t+1} \!\!= \!\!f(x^i_t, u^i_t),\;\!\!\! \forall t \in [T_s, T-1],\,\forall i \in [1, d] \\
        &h^i(x_t^i, u_t^i) \leq 0,\,  \forall t \in [T_s, T],\; 
        \forall i \in [1, d],\; 
    \end{aligned} 
\right.
\label{eq:risk_aware_motion_planning_problem}
\end{equation}
where $h^i(\cdot)$ encapsulates general inequality constraints for different branches, including control input bounds, collision avoidance constraints, and road boundary constraints.
The motion planning problem (RAMP) can be alternatively viewed as a two-player zero-sum game, where the first player seeks a comfortable and collision-free trajectory, while the second player acts as an adversary wishing to increase the total cost by redistributing the probability of each branch. In what follows, we propose an algorithm based on iLQR with augmented Lagrangian relaxation to solve the motion planning problem. 

\section{Efficient Risk-aware Motion planning}
In the subsequent developments, we draw inspiration from the methods in \cite{LiMultipolicy2023, LuHBSiA2020} to obtain a solution to (RAMP). 
Following \cite{LiMultipolicy2023}, we keep the probability vector $q$ fixed for each step and solve an approximated optimal control problem of (RAMP) via the iLQR scheme. 
We combine this method with a variant of HiBSA as proposed in  \cite{LuHBSiA2020}.
In particular, to improve convergence, we introduce a regularization term to the maximization problem, which perturbs the original projected gradient ascent step. 
As in \cite{LiMultipolicy2023}, we replace the projected gradient descent with a Newton-like step to further aid in convergence. Our proposed algorithmic scheme is detailed in Algorithm \ref{alg:al_ilqr_tree}. The vectors $\bar{\mathbf{u}}_{\text{ref}}$ and $\bar{\mathbf{x}}_{\text{ref}}$ denote the reference trajectories for the state and input, respectively. 

\subsection{Augmented Lagrangian iLQR tree}
 
\begin{algorithm}[t]
\SetAlgoLined
\KwIn{$x_0$, $\bar{\mathbf{x}}_\text{ref}$, $\bar{\mathbf{u}}_\text{ref}$, $p$}
\KwOut{$\bar{\mathbf{u}}^*$, $q$}
Initialize $\bar{\boldsymbol\lambda}^0$, $\mu^0$, $\bar{\mathbf{u}}^0 \leftarrow \bar{\mathbf{u}}_\text{ref}$, $k \leftarrow 0$\\
Perform rollout using $\bar{\mathbf{u}}_{\text{ref}}$ to obtain $\bar{\mathbf{x}}^{0}$ \\
\While{stopping criterion not satisfied}{
    \texttt{minmax\textunderscore iLQR\textunderscore tree}($x_0$, $\bar{\mathbf{x}}_\text{ref}$, $\bar{\mathbf{u}}^\text{k}$, $p$, $\bar{\boldsymbol\lambda}^k$, $\mu^k$)\;
    Update $\bar{\boldsymbol{\lambda}}^k$\;
    Update $\mu^k$\;
    $k \leftarrow k + 1$;
}
\caption{Risk-aware AL-iLQR tree} \label{alg:al_ilqr_tree}
\end{algorithm} 

To solve (RAMP) efficiently, we exploit the inherent sparse structure of the optimal control problem by leveraging ideas from dynamic programming  \cite{rawlings_model_2017}.    
This leads to an iLQR-based method, where the following subproblem is solved at each time step $t$:
\begin{equation} 
 \text{iLQR}_t^i\; \left\{
\begin{aligned}
    &\min_{u}\; Q^i_t(x, u, x^\prime)= V_t^i(x) \\
    &\text{ s.t.} \ \ \   x^\prime = f(x, u), \\
    &  \ \  \ \   \ \ \  h_t^i (x, u) \leq 0,
\end{aligned}
\right.
\label{eq:stage_optimization}
\end{equation}
where the $Q$-function $Q^i_t(\cdot)$ describes the cost incurred after applying the control input $u$ for branch $i$ at time step $t$, and $h_t^i(\cdot)$ represents general state and control input constraints.  
Based on dynamic programming, we compute the $Q$-functions for the trajectory tree as follows: 
\begin{align}
    Q^0_t(x, u) = \ell_t^0(x, u) + V^0_{t+1}\left(f(x, u)\right),\; t \in [0, T_s-2], \label{eq:value_function_shared} \\
    Q^0_{T_s-1}(x, u) = \ell_{T_s-1}^i(x, u) + \sum_{i=1}^{d} V^i_{T_s}\left(f(x, u)\right), \label{eq:value_function_branching}  \\
    Q^i_t(x, u) = \ell_t^i(x, u) + V^i_{t+1}\left(f(x, u)\right),\; t \in [T_s, T-1], \label{eq:value_function_branch}
\end{align}
where \eqref{eq:value_function_shared} and \eqref{eq:value_function_branch} are associated with the $Q$-functions of the shared branch and individual branches, respectively. Equation \eqref{eq:value_function_branching} represents the $Q$-function of the branching node, where the value functions for all branches at the next time step are summed up. 
Next, we consider the hard constraints by adding the augmented Lagrangian terms to the $Q$-functions:
\begin{multline*}
    Q_{a,t}^i (x, u, \lambda, \mu) = Q_{t}^i(x, u) + \lambda^{\top} h_t^i(x, u) \\ 
    + \frac{1}{2} h_t^i(x, u)^\top I_{\mu,t}^i  h_t^i(x, u),
\end{multline*}
where $\lambda$ is a vector of Lagrange multipliers, $\mu$ is a penalty weight. $I^i_{\mu,t}$ denotes a diagonal matrix that selects the active constraints:
\begin{align*}
    I^i_{\mu,t, mm} = \begin{cases}
    0 \quad \text{if}\; h_{t, m}^i(x, u) < 0\; \text{and}\;  \lambda_m = 0\\
    \mu \quad \text{otherwise},
    \end{cases}
\end{align*}
where,  $m$ is the index of the $m$th constraint.

By using augmented penalty terms, we now convert 
$\text{iLQR}_t^i$ into an unconstrained optimization problem. We then linearize the dynamics and approximate the $Q$-functions using second-order Taylor expansions. 
We employ the generalized Gauss-Newton Hessian approximation due to its ease of computation and the theoretical guarantee that the approximated Hessian is always positive semi-definite \cite{schraudolph_fast_2002}.
The Newton descent direction at time step $t$ can then be obtained by minimizing the approximated version $\hat{Q}^i_{a,t}$ of the augmented Langrangian $Q$-function: 
\begin{align}
    \min_{\delta u}\; \hat{Q}_{a,t}^i (\delta x, \delta u). \label{eq:stage_optimization_approx}
\end{align}
Since $\hat{Q}_{a,t}^i$ is in quadratic form, we can derive an affine control law of the form $\delta u = K_t^i \delta x + d_t^i$ \cite{HowellALTRO2019}. 
The backward pass involves solving \eqref{eq:stage_optimization_approx} from the leaf tree nodes to the root node.
It is worth noting that the computation from the leaf nodes to the branching node can be performed in parallel, thus improving the computation speed significantly.
After the backward pass, we conduct a forward rollout using the nonlinear dynamics to obtain the updated trajectory tree. 
A standard backtracking line search \cite{Nocedal_numerical_2006} is performed to guarantee a sufficient decrease in the trajectory cost.
Similarly to \cite{TassaControliLQR2014}, we introduce a regularization term to guide the Newton direction towards the gradient descent direction in case the line search fails.

\subsection{Projected gradient ascent with regularization}
We note that the risk-aware motion planning in \eqref{eq:risk_aware_motion_planning_problem} is a nonconvex-concave problem. 
For this class of problems, applying the GDA method directly might result in oscillations \cite{RazaviyaynMinmax2020}.
Therefore, motivated by \cite{LuHBSiA2020}, 
we render the original cost function \eqref{eq:CVaR_cost} strongly concave with respect to $q$ by introducing a quadratic regularization term:
\begin{multline}
    J_{\text{R}}(x_0, \bar{\mathbf{u}}) := J^0(x_0, \bar{\mathbf{u}}^0) + \\ 
    \max_{q \in \mathcal{A}_\alpha(p)}\; \sum_{i=1}^d q^i J^i(x_{T_s}, \bar{\mathbf{u}}^i) - \frac{\rho}{2} q^{i2}, \label{eq:regularized_max} 
\end{multline}
where $\rho > 0$ is the regularization weight.
A high value of $\rho$ drives the probability vector $q$ towards the centroid of the simplex.
However, we note that this additional regularization term modifies the saddle point of the original problem. 
To mitigate its impact on the saddle point, we adopt a diminishing regularization weight given by $\rho^k = \rho^0 / (k+1)$.
We then perform a projected gradient ascent to approximately solve \eqref{eq:regularized_max} at iteration $k$:
\begin{align}
    q^{k+1} \leftarrow \mathrm{proj}_{\mathcal{A}_\alpha(p)} \left( (1 - \gamma\rho^k)q^k + \gamma \bar{J}(x_{T_s}, \bar{\mathbf{u}}) \right), \label{eq:q_update}
\end{align}
where $\bar{J}(\cdot)=(J^i(\cdot))_{i \in [1,d]}$ is the collection of all branch costs except for the shared branch, $\mathrm{proj}(\cdot)$ is the projection operator, and $\gamma > 0$ is the step size.  
We compute the projection onto the ambiguity set as follows \cite[Chapter 6.4.3]{Beck2017}:
\begin{align*}
    \mathrm{proj}_{\mathcal{A}_\alpha(p)}(q) = \mathrm{proj}_{\mathrm{box}[0,\frac{p}{\alpha}]} (q - \phi^* \boldsymbol{1}_{d}),
\end{align*}
where $\mathrm{box}[0,\frac{p}{\alpha}] := \{ q \in \mathbb{R}^d \mid 0 \leq q^i \leq \frac{p^i}{\alpha},  \forall \ i \in [1,d] \}$, $\boldsymbol{1}_{d} = (1, 1, \dots, 1) \in \mathbb{R}^d$ represents a vector with all elements being $0$, 
and $\phi^*$ is a root of the equation:
\begin{align*}
    m(\phi) := \boldsymbol{1}_d^\top \mathrm{proj}_{\mathrm{box}[0,\frac{p}{\alpha}]} (q - \phi \boldsymbol{1}_d) - 1 = 0. 
\end{align*}
According to \cite[Chapter 6.4.3]{Beck2017}, $m(\phi)$ is a non-increasing function. Thus, its root can be efficiently computed through the bisection method.  

\subsection{Detailed formulation}
\subsubsection{Vehicle modeling} We model the vehicle as a kinematic bicycle with the state vector $\tilde{x} := (p_{x}, p_{y}, \theta, v)$ and the control input vector $u := (a, \delta)$, where
$(p_{x}, p_{y})$, $\theta$, and $v$ represent the rear-axle position, heading angle, and speed of the vehicle, respectively; 
$a$ and $\delta$ are the acceleration and steering angle. 
In \eqref{eq:risk_aware_motion_planning_problem}, the stage cost only depends on the current state and control input. 
Therefore, to penalize the rate of change of the control inputs, we augment the original state by concatenating it with the previous control input, resulting in $x_k := [\tilde{x}_k^\top,\; u_{k-1}^\top]^\top$. Additionally, we ensure feasibility of physical quantities such as speed, acceleration, and steering angle by constraining them within lower and upper limits.
\begin{algorithm}[t] \label{alg:ilqr_tree}
\SetAlgoLined
\KwIn{$x_0$, $\bar{\mathbf{x}}_\text{ref}$, $\bar{\mathbf{u}}^\text{0}$, $p$, $\bar{\boldsymbol\lambda}$, $\mu$}
\KwOut{$\bar{\mathbf{u}}^*$, $q$}
Perform rollout using $\bar{\mathbf{u}}^{0}$ to obtain $\bar{\mathbf{x}}^{0}$ \\
\While{stopping criterion not satisfied}{
    Approximate $Q$-functions\;
    Compute backward pass through Riccati equation\;
    Compute forward pass, including rollout and line search\;
    \uIf{line search failed}{
        Add regularization terms to Hessian matrices\;
    }
    \Else{
        Update the probability vector $q$ using \eqref{eq:q_update}\;
    }    
}
\caption{minmax\textunderscore iLQR\textunderscore tree}
\end{algorithm}
\subsubsection{Cost function}
The motion planner is designed to track the reference trajectory tree, maximize the driving comfort level, and keep a safe distance from surrounding vehicles.
To account for the driving comfort, we penalize the control input and its rate of change.
Moreover, we design the safety cost $\ell: \mathbb{R}^2 \rightarrow \mathbb{R}$ between the ego vehicle and the other vehicle at positions $p^{\text{EV}}$ and $p^{\text{SV}}$, respectively, as follows:
\begin{align*}
    \ell_{\text{saf}}(p^{\text{EV}}, p^{\text{SV}}) = 
    \begin{cases}
        (d - d_{\text{prox}})^2 &\text{if}\; d < d_{\text{prox}} \\
        0 &\text{otherwise},
    \end{cases}
\end{align*}
where $d$ is the Euclidean distance between the center positions of the ego vehicle and the other vehicle,  
and $d_{\text{prox}}$ represents the threshold distance.

\subsubsection{Safety constraints}
The distance between two vehicles is a nonsmooth function, which poses numerical challenges to the optimization algorithm. 
To account for the lack of smoothness, we use instead an overapproximation of the shape of each vehicle, which comprises the union of a collection of linked circles \cite{WangInteraction2023}. 
We then compute the smooth collision avoidance constraints as follows:
\begin{align*}
    (r^{\text{EV}} + r^{\text{SV}})^2 - \left\Vert c^i(x^{\text{EV}}) - c^j(x^{\text{SV}}) \right\Vert_2^2 \leq 0, \\i \in [1, n_c^{\text{EV}}],\; j \in [1, n_c^{\text{SV}}],
\end{align*}
where $r^{\text{EV}}$ and $r^{\text{SV}}$ represent the safety circles' radii, corresponding to the ego vehicle and the other vehicle,  respectively, $c^i: \mathbb{R}^{n_x} \rightarrow \mathbb{R}^2$ is a function that computes the center of each circle, 
and $n_c^{\text{EV}}$, $n_c^{\text{SV}}$ denote the number of circles used to approximate the vehicle footprint. \par 
Finally, enforcing boundary constraints for curved roads can be challenging for motion planning in Cartesian coordinates. 
Following \cite{liniger_optimization-based_2015}, we approximate the road boundary constraints by constructing a safe driving corridor along the reference trajectory tree provided by the behavior planner.

\section{Numerical study}

\subsection{Simulation setup}
\begin{figure}[t]
    \centering
    \begin{subfigure}[b]{0.48\columnwidth}
        \includegraphics[width=\textwidth]{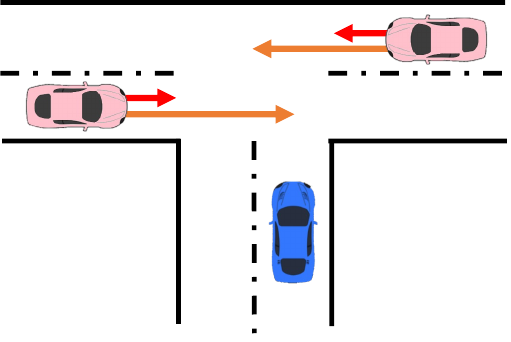}
        \caption{Test scenario 1 (TS1)}
        \label{fig:ts_1}
    \end{subfigure}
    \hfill
    \begin{subfigure}[b]{0.48\columnwidth}
        \includegraphics[width=\textwidth]{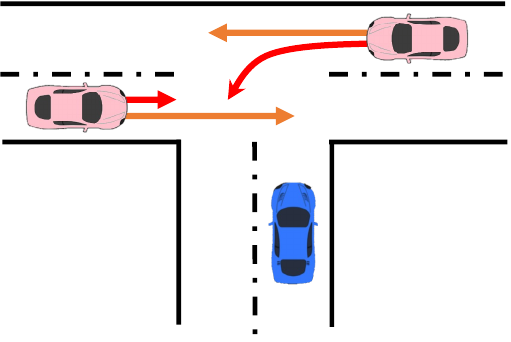}
        \caption{Test scenario 2 (TS2)}
        \label{fig:ts_2}
    \end{subfigure}
    \caption{Test scenarios. The ego vehicle (in blue) intends to turn left. (a) Both surrounding vehicles have two potential behavior modes: ``Yield'' and ``Assert'', represented by the red and orange arrows, respectively. (b) The upper vehicle exhibits different behavior modes: ``TurnLeft'' and ``GoStraight''.}
    \label{fig:ts}
\end{figure}

We test our method in two distinct unsignalized intersection-crossing scenarios illustrated in Fig. \ref{fig:ts}. 
In both scenarios, the ego vehicle (in blue) intends to turn left, while two surrounding vehicles are present with different driving intentions per scenario. 
Specifically, in the first scenario (TS1), both surrounding vehicles have two potential behavior modes: ``Yield'' and ``Assert'', with their corresponding motion plans represented by the red and orange arrows, respectively. 
In the second scenario (TS2), the vehicle on the right exhibits two different behavior modes: ``TurnLeft'' and ``GoStraight''.
Since, for each case study, two possible behavior modes are considered for each surrounding vehicle, we have in total four different combinations of behavior modes or, equivalently, four separate branches in the trajectory tree. \par 

Note that the choice of the initial guess of the trajectory tree can significantly influence the convergence speed or in some cases hinder the convergence of the motion planner. 
As such, the initial guess should be appropriately selected.
In our setup, we adopt a simple sampling-based behavior planner \cite{ZhangGame2023, DingEPSILON2022}.  
Specifically, we control the longitudinal and lateral motion of the ego vehicle via a PD controller and a pure pursuit controller, respectively, and forward simulate its motion under different desired speeds to obtain a set of trajectories. 
For each joint behavior mode, we select the best trajectory from the trajectory set based on certain user-defined criteria.

Our motion planner operates at 10 Hz with a
discretization step of \SI{0.1}{\second} and a planning horizon of \SI{5}{\second}. 
The number of shared nodes $T_s$ is set to $5$.
We conduct all simulations on a laptop with a \SI{2.30}{\giga\hertz} Intel Core i7-11800H processor and \SI{16}{\giga\byte} RAM.
The motion planning algorithm is coded in C++.

\subsection{Numerical convergence results}
We run open-loop Monte Carlo simulations to empirically validate the convergence of the proposed motion planner under $500$ different initial states of the ego vehicle.
To obtain these states, we perturb the nominal initial state by $\SI{\pm 3}{\meter}$ for the longitudinal position, $\SI{\pm 1}{\meter}$ for the lateral position, $\SI{\pm10}{\percent}$ for the longitudinal speed.
Additionally, we set $p^i = 0.25$ for all $i \in [1, d]$ and $\alpha = 0.6$ when constructing the ambiguity set.
We compare our motion planner with MARC \cite{LiMultipolicy2023} and Dual MPC \cite{ChenMultiModal2022}.
We implement Dual MPC using the IPOPT \cite{wachter_implementation_2006} interface provided by CasADi \cite{andersson_casadi_2019} with MA57 \cite{duff_ma57---code_2004}  as the linear solver for enhanced performance.
The statistical results on convergence and average computation time are presented in Table \ref{tb:st_res}. 
Our method achieves successful convergence in the majority of cases. However, when employing MARC, we observe that in certain cases the values of the probability vector $q$ oscillate between two vertices of the ambiguity set.
A potential reason is that the solver of the linear program often outputs a vertex as the optimal solution, but the saddle point might be a point on a facet of the ambiguity set that is not necessarily a vertex.
Dual MPC has the highest computational cost since additional decision variables are introduced and the original nonconvex cost becomes a part of hard constraints after dualization. 
Consequently, such reformulation is often not well-suited for efficient computation.

\begin{figure}
    \centering
    \includegraphics[width=\columnwidth]{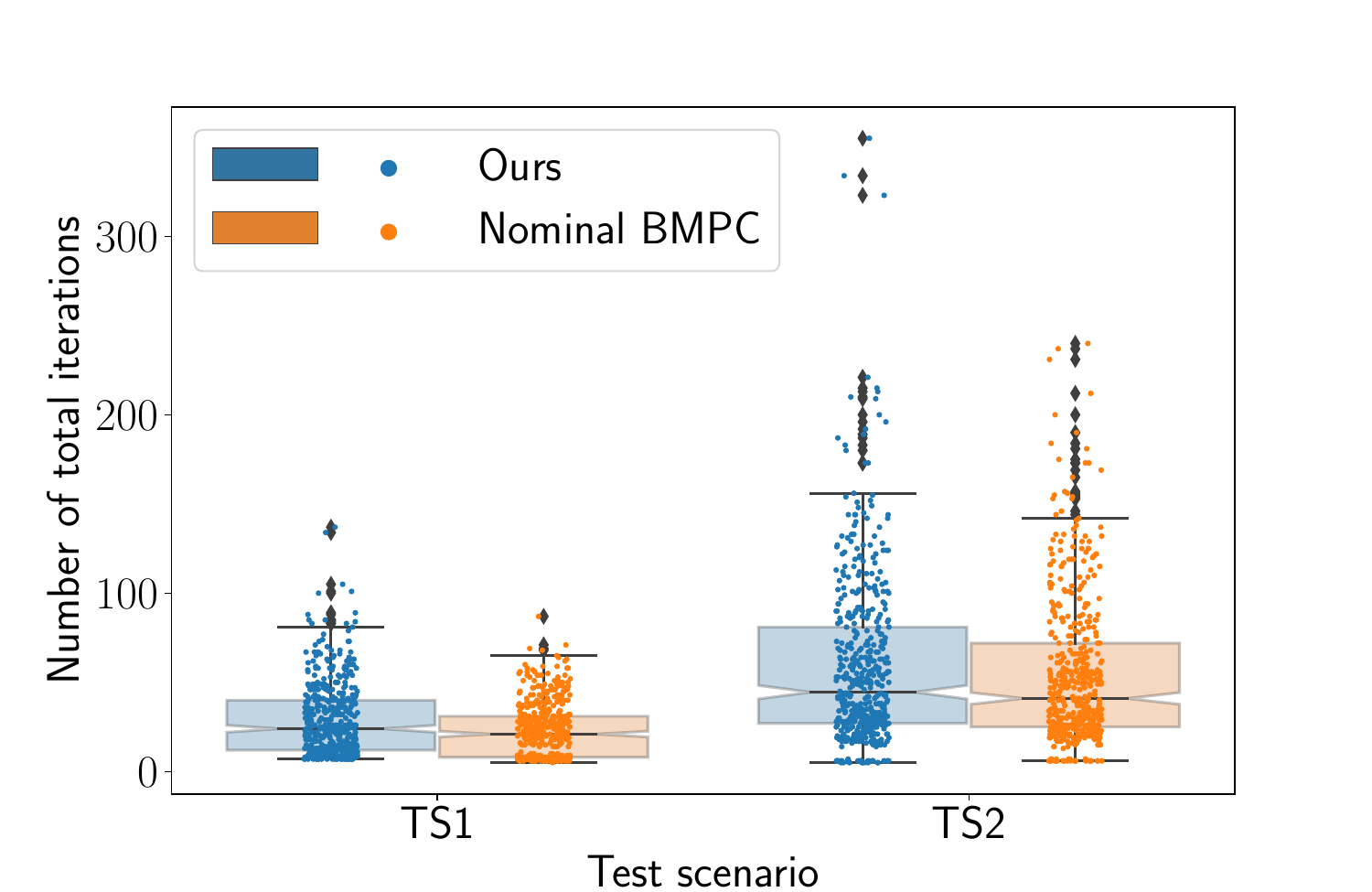}
    \caption{Box-plots obtained from 500 Monte Carlo simulations: The total number of iterations to solve our risk-aware branch MPC problem (blue dots) is for some cases larger compared to those required to solve the nominal branch MPC problem (orange) in both test scenarios (TS1) and (TS2). The difference is due to the presence of gradually decaying oscillations in the probability update for the case of risk-aware MPC.}
    \label{fig:num_iter}
\end{figure}

Fig. \ref{fig:num_iter} illustrates that the number of total iterations required to solve a risk-aware branch MPC problem is slightly larger than that required for the nominal branch MPC problem.
Simulation studies indicate that the difference in the number of required iterations is due to the presence of gradually decaying oscillations in the probability update.
Such an update is not present in the case of the nominal branch MPC. 
Note that such extra computational overheads are still acceptable from a practical point of view, since the projection in \eqref{eq:q_update} can be efficiently computed in just a few microseconds, and the average computation time of our method is below \SI{100}{ms}. 
Thus, the proposed planner is well-suited for real-time motion planning after code optimization.

\setlength{\tabcolsep}{4.5pt} 
\begin{table}[t]
  \begin{center}
    \caption{Statistical Results}
    \label{tb:st_res}
    \begin{tabular}{@{}ccccc@{}}
    \toprule
    Test Scenario & Metric & Ours & MARC \cite{LiMultipolicy2023} & Dual MPC \cite{ChenMultiModal2022}   \\ \midrule
    \begin{tabular}[c]{@{}c@{}} TS1\end{tabular}   & \begin{tabular}[c]{@{}c@{}} Convergence \\ Comp. time (\unit{\milli\second}) \end{tabular} & \begin{tabular}[c]{@{}c@{}}
    \cellcolor{lightgray}\SI{100}{\percent}\\ \cellcolor{lightgray}25.9  \end{tabular}     & \begin{tabular}[c]{@{}c@{}} \SI{90.6}{\percent} \\ 36.9 \end{tabular}      &\begin{tabular}[c]{@{}c@{}} \SI{100}{\percent}\\ 1300.8 \end{tabular}      \\
    \begin{tabular}[c]{@{}c@{}} TS2 \end{tabular} & \begin{tabular}[c]{@{}c@{}} Convergence \\ Comp. time (\unit{\milli\second}) \end{tabular} & \begin{tabular}[c]{@{}c@{}} 99.8 \unit{\percent}\\ \cellcolor{lightgray}53.1 \end{tabular} & \begin{tabular}[c]{@{}c@{}} 95.2 \unit{\percent}\\ 59.7 \end{tabular} & \begin{tabular}[c]{@{}c@{}} \cellcolor{lightgray}\SI{100}{\percent}\\ 1520.7 \end{tabular} \\ \bottomrule
    \end{tabular}
  \end{center}
\end{table}

\subsection{Closed-loop trajectories}
We now compare the closed-loop performance between the risk-aware branch MPC and the nominal branch MPC for the scenario (TS1). 
\begin{figure}[!t]
     \centering
     \begin{subfigure}[b]{0.9\columnwidth}
         \centering
         \includegraphics[width=\textwidth]{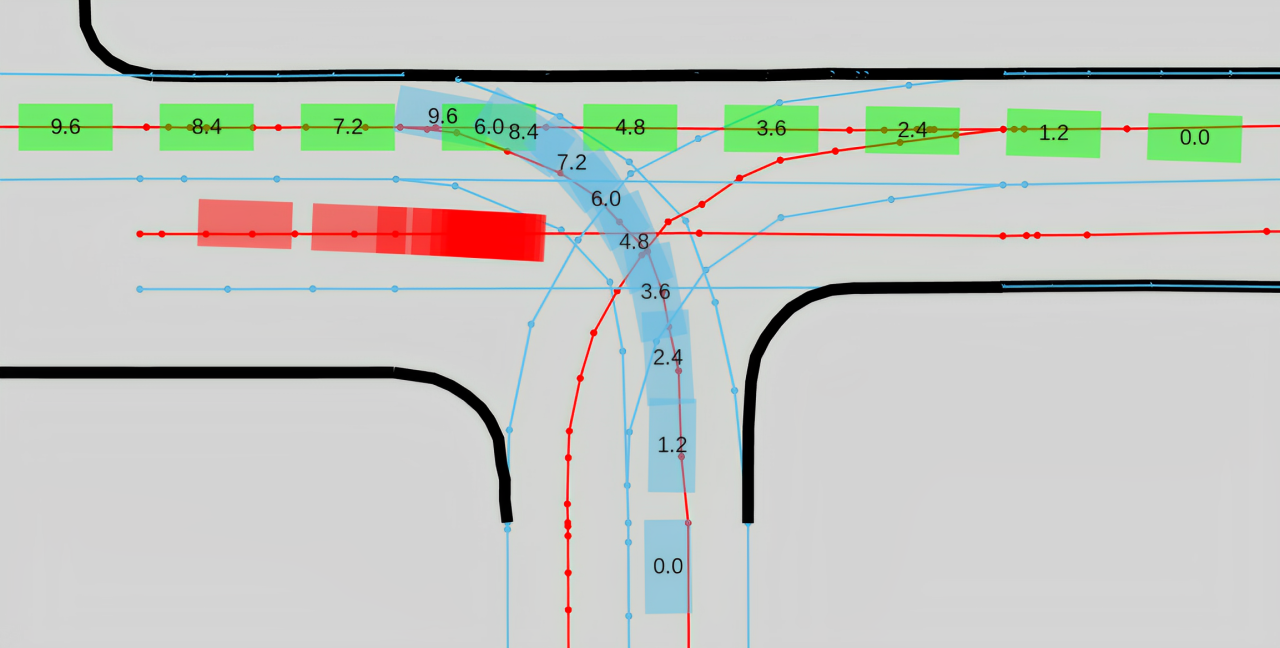}
         \caption{Closed-loop trajectories of risk-aware motion planning.}
         \label{fig:cl_trajs}
     \end{subfigure}
     \hfill
     \begin{subfigure}[b]{1.0\columnwidth}
         \centering
         \includegraphics[width=\textwidth]{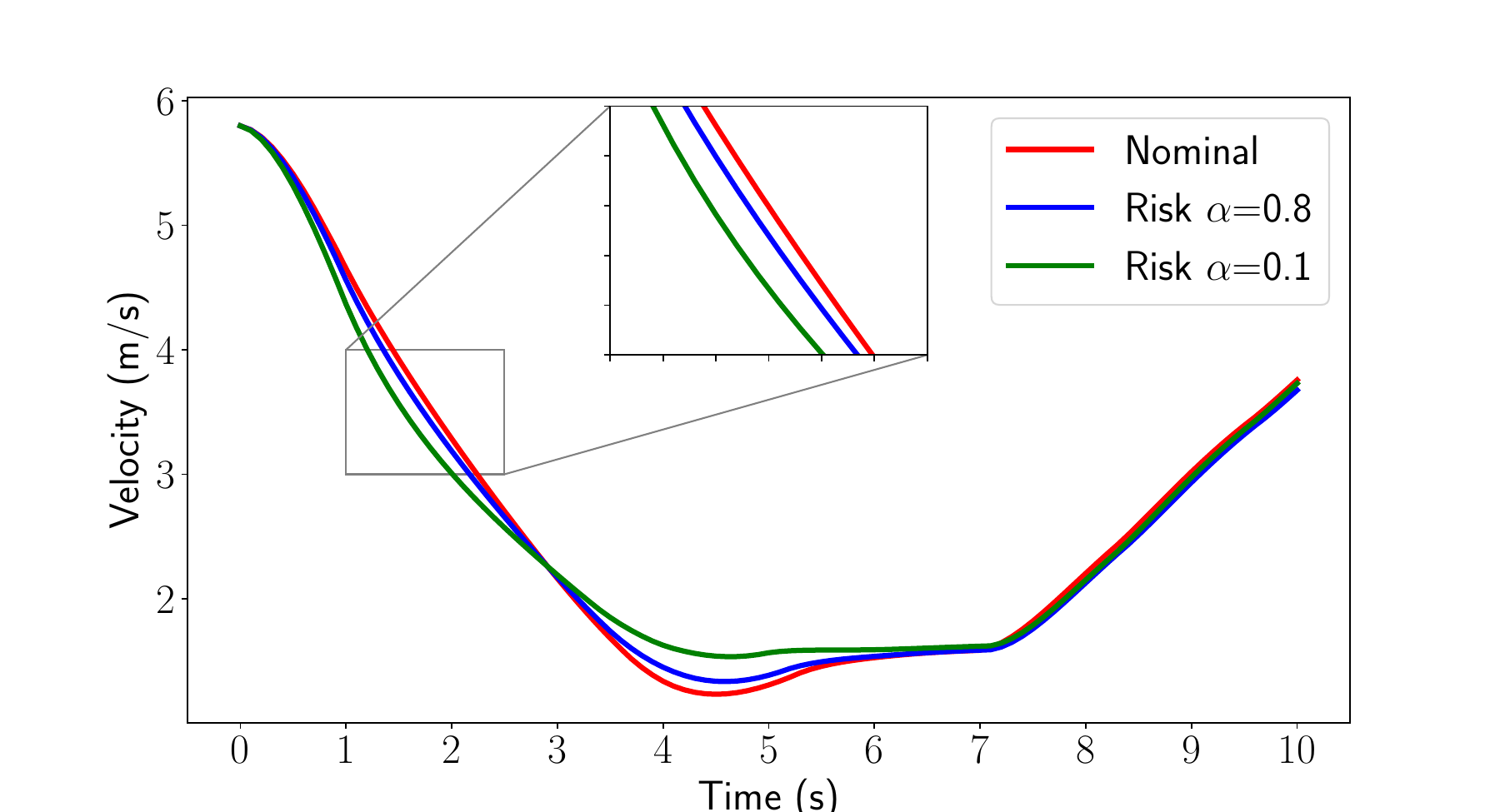}
         \caption{Velocity profiles generated by different planners.}
         \label{fig:vel_profile}
     \end{subfigure}
    
        \caption{(a) Closed-loop trajectories of risk-aware motion planning:  The actual behavior of the red vehicle is to yield, while the green one keeps a constant velocity. Initially, since the ego vehicle (in blue) is unaware of the true intentions of the surrounding vehicles, it slows down and merges behind the green vehicle to mitigate the risk of the red vehicle not yielding.
        (b) Velocity profile comparison among risk-aware motion planners for two values of the risk parameter $\alpha=0.1$ and $\alpha=0.8$ and the nominal motion planner.} 
        \label{fig:cl_sim}
\end{figure}
Specifically, we assume that the ego vehicle does not fully know the behavior modes of the surrounding vehicles before $t_a=\SI{1}{\second}$.  
In practice, the ego vehicle can identify the intentions of the surrounding vehicles after a certain period. 
Considering this, we assume that the ego vehicle has complete knowledge of these intentions after $t_a=\SI{1}{\second}$. 
Simulation studies indicate that, as expected, an increase in $t_a$, results in gradually more conservative behavior.
As the risk-aware planner accounts for the risk against different behavior mode realizations before $t_a$, its trajectory differs from that of the nominal motion planner.
Note that even after $t_a$, when the ego vehicle knows the intentions of the surrounding vehicles, the trajectory of the risk-aware and nominal planners are still different since the initial differences in the trajectory affect their future evolution.
The risk-aware motion plan is illustrated in Fig. \ref{fig:cl_trajs}. 
The number on the snapshot of each vehicle denotes the simulation timestamp.
At first, the ego vehicle does not know whether the red vehicle will yield or continue its straight trajectory. To account for this risk, the ego vehicle chooses to slow down and merge behind the green vehicle. \par 
Fig. \ref{fig:vel_profile} shows the velocity profile of the ego vehicle for the nominal motion planner and the risk-aware motion planner for two distinct values of the risk parameter $\alpha \in [0,1]$. 
We remind the reader that smaller values of $\alpha$ imply the presence of more ambiguity concerning the probabilistic estimates of the behavior modes of the red and green vehicles. 
We observe that for up to $t=\SI{3}{\second}$,  the velocity of the risk-aware motion is slower than the nominal one, as the risk-aware approach considers the risk of the red vehicle not yielding. 
After $t=\SI{3}{\second}$ the velocity increases compared to the nominal motion planner, and the profiles coincide after $t=\SI{7}{\second}$. 
Finally, we note that for $\alpha=0.1$, the motion planner is, as expected, more risk-aware of the behavior of the red vehicle, while for  $\alpha=0.8$, its behavior is closer to the nominal one, since ambiguity is smaller.

\section{Conclusion}
Risk-aware branch model predictive control is applicable to motion planning for automated vehicles subject to behavioral uncertainty. 
However, general-purpose numerical solvers are currently not fast enough to solve the considered motion planning problem within the given sampling time. 
Our method shows high potential in closing this computational gap and paves the way towards more efficient real-time motion planning.
In future work, we aim at designing an efficient interaction-aware behavior planner for intersection-crossing scenarios. 
Our final goal is to develop a general motion planning framework by integrating an advanced behavior planner with the proposed risk-aware motion planner.  

\printbibliography
\flushend 


\end{document}